\documentclass[aps,prb,twocolumn,groupedaddress]{revtex4-2}

\usepackage{amssymb}
\usepackage{amsmath}
\usepackage{bm}
\usepackage{graphicx}
\usepackage{dcolumn}
\usepackage{longtable}
\usepackage{color}
\usepackage{multirow}

\begin{document}


\title{Field-induced magnetic structures in the chiral polar antiferromagnet Ni$_2$InSbO$_6$}

\author{Y.~Ihara}
\email{yihara@phys.sci.hokudai.ac.jp}
\author{R.~Hiyoshi}
\author{M.~Shimohashi}
\author{R.~Kumar}
\affiliation{Department of Physics, Faculty of Science, Hokkaido University, Sapporo 060-0810, Japan}
\author{T.~Sasaki}
\author{M.~Hirata}
\altaffiliation[present address]{MPA-Q, Los Alamos National Laboratory, Los Alamos, 87545 New Mexico, USA}
\affiliation{Institute for Materials Research, Tohoku University, Sendai, 980-8577, Japan}
\author{Y.~Araki}
\author{Y.~Tokunaga}
\affiliation{Department of Advanced Materials Science, University of Tokyo, Kashiwa 277-8561, Japan}
\author{T.~Arima}
\affiliation{Department of Advanced Materials Science, University of Tokyo, Kashiwa 277-8561, Japan}
\affiliation{RIKEN Center for Emergent Matter Science (CEMS), Wako 351-0198, Japan}
\date{\today}

\begin{abstract}
We have performed $^{115}$In-NMR spectroscopy for Ni$_{2}$InSbO$_6$ with corundum-related crystal structure to reveal magnetic structures that develop in high magnetic fields. 
At low fields Ni$_{2}$InSbO$_6$ shows a helical magnetic order with a long wavelength because of its chiral and polar crystal structure. 
The field-induced magnetic state was not investigated by microscopic experiment because 
an extremely high magnetic field is required to modify the antiferromagnetically coupled helical structure. 
From the analysis of our $^{115}$In-NMR spectra obtained at high magnetic fields, we confirm that the canted antiferromagnetic structure appears in fields applied in the $[110]$ direction 
and the propagation vector of magnetic helix is rotated toward the field direction for fields in the $[001]$ direction. 
We discuss the effect of magnetic field that modifies the magnetic structure of an antiferromagnetic chiral magnet.
\end{abstract} 

\maketitle

\section{Introduction}

The correlation between the space-inversion breaking and magnetism, leading linear magnetoelectric (ME) effect between the magnetic (electric) field and electric (magnetic) polarization, 
has long been attracting much attention because of their fundamental interest and technological applications.  
The antiferromagnetic (AFM) order in corundum-type Cr$_2$O$_3$ simultaneously breaks the space-inversion and time-reversal symmetries, 
which triggers the linear ME effect. \cite{astrov-SPJ11, folen-PRL6}
Dzyaloshinskii and Moriya succeeded in elucidating the origin of a weak spontaneous magnetization in corundum type Fe$_2$O$_3$ \cite{dzyaloshinsky-JPCS4, moriya-PR120}
by introducing the antisymmetric exchange term, 
which sometimes induces a helical magnetic order with a long modulation period. 
The magnetic helix can be further modulated by the application of a magnetic field. 
The response to the external magnetic field should be explored to understand and control the ME effect potentially applicable to novel devices. 

In chiral magnets, the application of magnetic fields gradually modifies the magnetic helix at zero field  
into a magnetic chiral soliton lattice in CrNb$_3$S$_6$ \cite{togawa-PRL108} 
and the nanometric magnetic swirling object termed skyrmion in MnSi assisted by the thermal fluctuations. \cite{muhlbauer-Science323}
In contrast to these prototypal chiral magnets with ferromagnetic (FM) symmetric exchange interactions, 
the magnetic field effects on the ME materials with AFM interaction have not been investigated in detail, 
because the critical fields required to modify the AFM structure are much higher than those in the cases of FM chiral magnets. 
In the AFM chiral magnets, thus, it is not obvious if the application of a magnetic field can actually modify the magnetic helix. 
As the theoretical prediction of the field effect on the AFM structure is difficult when nearly isotropic magnetic moments reside in the chiral and polar crystal structure, 
it is essential to experimentally reveal a change in AFM structure by magnetic fields for unveiling the origin of the ME effect in AFM chiral magnets. 

\begin{figure}[tbp]
\begin{center}
\includegraphics[width=8.5cm]{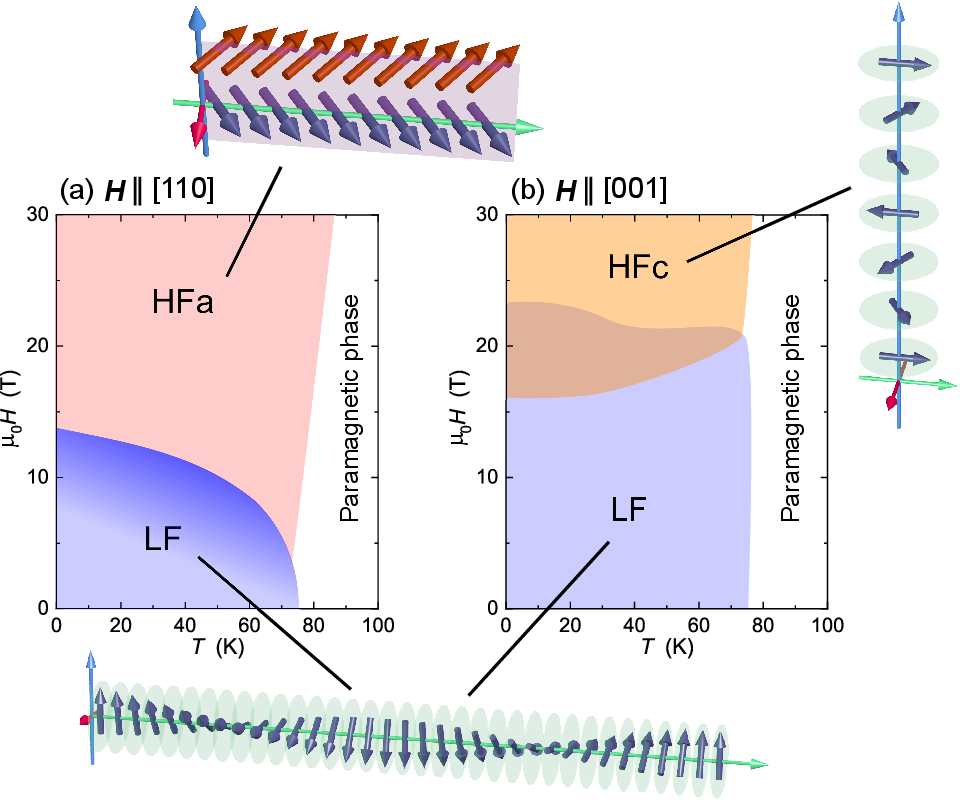}
\end{center}
\caption{
Magnetic phase diagram of Ni$_2$InSbO$_6$ for (a) $\bm{H} \parallel [110]$ and (b) $\bm{H} \parallel [001]$ directions. \cite{araki-PRB102}
The magnetic structure at small magnetic fields is a proper helical with long wavelength (LF phase). \cite{ivanov-CM25}
(a) In the fields along $[110]$, soliton excitations are introduced near the metamagnetic fields of $B_m^a=14 $ T. 
Above $B_m^a$, canted AFM state (HFa phase) is stabilized. 
(b) In the fields along $[001]$, the propagation vector of helical structure flops to $[001]$ direction above of $B_m^c=19.2 $ T (HFc phase).
Hysteresis between LF and HFc phases was observed in the magnetization and electric polarization measured in pulsed magnetic fields. 
}
\label{fig1}
\end{figure}

As represented by Cr$_2$O$_3$ and Fe$_2$O$_3$, magnetic oxide compounds of corundum-related structures provide a good playground for investigating the AFM state emerging on the space-inversion-symmetry broken crystal structure. 
Particularly, if the cation sites are periodically occupied by more than one element, the cation ordering may cause a further lowering of symmetry. 
Ni$_2$InSbO$_6$ is such a compound, which crystallizes in a hettotype structure. \cite{ivanov-CM25}
The cation sites of corundum form honeycomb layers perpendicular to the three-fold axis. 
In Ni$_2$InSbO$_6$, Ni-In and Ni-Sb honeycomb layers are alternately stacked. 
Ni and In or Sb are alternately arranged in each layer as in hexagonal boron nitride. 
As a consequence, the centers of inversion and glide mirrors in the corundum (space group $R\bar{3}c$) are completely broken, 
and the space group becomes chiral and polar $R3$. 
As is expected, helical magnetic order with an in-plane small wave vector (LF phase) has been confirmed below $T_N = 78$ K by neutron studies at zero magnetic field. \cite{ivanov-CM25}
It is proposed that the spin spiral plane should be perpendicular to the wave vector, which implies weak magnetic anisotropy in the Ni$^{2+}$ moments. \cite{araki-PRB102} 
The magnetic transition is also accompanied by a kink anomaly in the temperature dependence of electric polarization, 
which suggests that the magnetic ordering on the space-inversion-broken crystal structure has introduced the ME response. \cite{araki-PRB102}

Moreover, the measurements of magnetization and electric polarization in a pulsed high magnetic field have revealed metamagnetic transitions with a fairly large ME effect. \cite{araki-PRB102}
At the critical fields $B_m^{a} \simeq 14$ T for $\bm{H} \parallel [110]$ and $B_m^{c} \simeq 19$ T for $\bm{H} \parallel [001]$, 
electric polarization shows an anomaly, which suggests the modification of ME coupling tensor by the field-induced magnetic structure transition. 
Various magnetic structures can be stabilized by magnetic fields in Ni$_2$InSbO$_6$ with small magnetic anisotropy and DM and AFM interactions. 
Previous study suggested a canted AFM and a Q-flopped magnetic structures for the high-field phases 
in $\bm{H} \parallel [110]$ (HFa phase) and $\bm{H} \parallel [001]$ (HFc phase), respectively, as summarized in Fig.~\ref{fig1}. \cite{araki-PRB102} 
In order to solve the magnetic structure in high magnetic fields microscopic experiments, worthy of addressing the relative orientations of magnetic moments, are desirable. 
It is, however, not easy to perform a neutron diffraction and a resonant x-ray scattering measurements in a pulsed high magnetic field. \cite{ohoyama-JMMM310, nojiri-PRL106, duc-RSI89} 
We must utilize a complementary experimental technique that allows us to probe the microscopic magnetic structure and can be performed in very high magnetic fields. 
Therefore, in this study, we measured the nuclear magnetic resonance (NMR) spectrum in fields up to 24 T to analyze the field-induced magnetic structures in Ni$_2$InSbO$_6$. 
From our $^{115}$In-NMR spectrum measurement we observed the internal fields at In sites, 
which are generated by the ordered Ni$^{2+}$ moments, and thus their strength depends on the local spin configurations around the In sites. 
We estimate the microscopic parameters that describe the magnetic coupling strength between Ni$^{2+}$ moments and In nuclear spins from the 
analyses in the high-temperature paramagnetic state and LF phase. 
Using these parameters, we determined the magnetic structures in the field-induced magnetic state. 

\section{Experimental }

Single crystals of Ni$_{2}$InSbO$_{6}$ were grown with the chemical vapor transport method \cite{weil-CRT49} and the dimension of the sample used for NMR measurement is $1\times1\times0.5$ mm$^3$. 
We confirmed by the Laue photography that a domain that has the $a$ axis rotated by 30 degrees about the $c$ axis is absent. 
In the $R3$ space group without inversion and mirror symmetries, chiral and polar domains exist inevitably. 
The domain formation does not affect the analyses of NMR spectra, because we focus only on the local magnetic fields around the In sites. 
A high-field NMR spectrum measurement was performed with a cryogen-free superconducting magnet at the Institute for Materials Research (IMR), Tohoku University. \cite{awaji-IEEE24}
Field-sweep NMR spectra were reconstructed from the transient Fourie transform (FT) spectra obtained during the field sweep at a constant sweep rate. \cite{clark-RSI66}
Broad NMR spectra in the ordered state were measured at several frequencies to follow the field variation of spectral shape. 
Field-dependence of NMR spectra was measured without changing the radio-frequency (rf) coil for the rf tank circuit.  
As all the NMR spectra in one field orientation were measured {\it in situ}, 
we can exclude the effect of change in the field orientation for the field-induced modification of spectral structure. 
For a high-temperature measurement at a fixed field of 13.0248 T, 
we measured only a sharp central peak, which allows us to cover full spectrum by the frequency window (typically $\sim 500$ kHz) for a single FT spectrum.

\begin{figure}[tbp]
\begin{center}
\includegraphics[width=7.5cm]{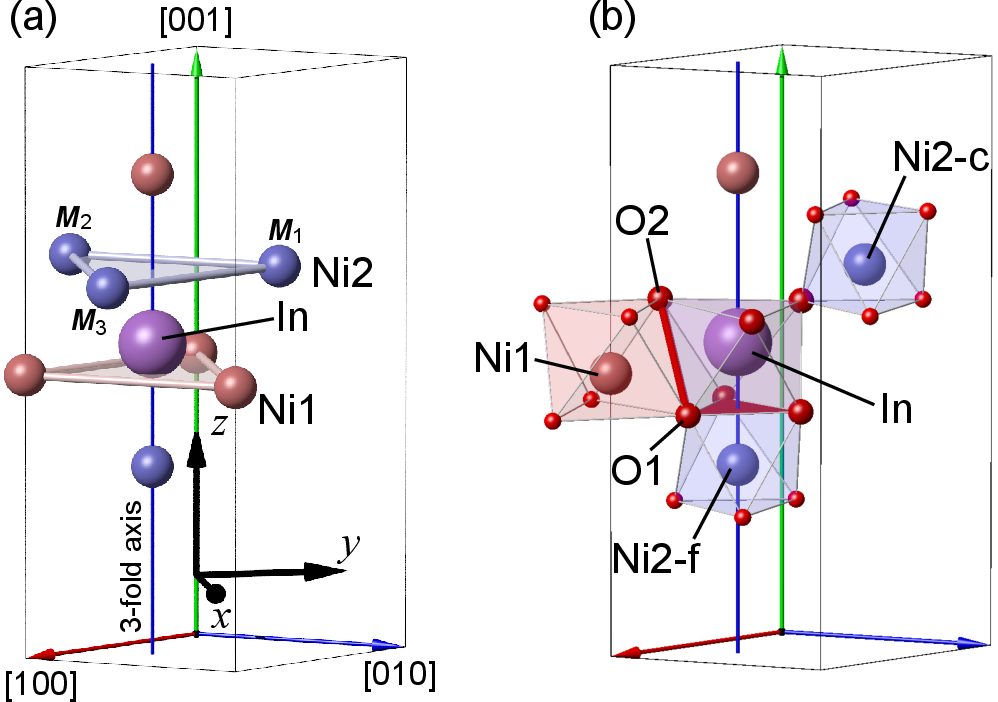}
\end{center}
\caption{
Ni sites around the NMR-target In site on the three-fold axis. 
(a) Eight Ni sites within a distance of $4$ \AA. 
The cartesian coordinate system for the hyperfine coupling tensor is indicated as $xyz$ axes. 
(b) NiO$_6$ octahedra around the In site. 
Ni$^{2+}$ moments are coupled with In nuclear spin through the Ni-O-In exchange paths. 
Ni2-c, Ni1, and Ni2-f sites have one, two and three exchange paths, respectively. 
}
\label{fig2}
\end{figure}

\section{results and discussions}
\subsection{NMR parameters}

NMR spectroscopy is a useful microscopic technique to investigate the magnetic structure in magnetic fields. 
In the magnetically ordered state the ordered moments create internal fields at the target nuclear sites, which shift the peak positions of the NMR spectra. 
We compute the internal fields for several model magnetic structures, and compare the result with those estimated from the experimentally obtained NMR spectra. 
The nuclear spins are magnetically coupled with the ordered moments through the dipole-dipole and hyperfine interactions. 
The dipole fields can be directly calculated from the crystal and magnetic structures. 
To estimate the hyperfine fields we need to determine experimentally the hyperfine coupling strength $A_{\rm hf}$ between Ni$^{2+}$ moments and In nuclear spins. 
In general the hyperfine coupling constant is written in a tensor form as
\begin{align}
\widehat{ A}_{\rm hf} = \left( 
\begin{array}{ccc}
A_{xx} & A_{xy} & A_{xz} \\
A_{yx} & A_{yy} & A_{yz} \\
A_{zx} & A_{zy} & A_{zz} \\
\end{array}
\right),
\end{align}
where $z$ and $x$ directions are parallel to crystallographic $[001]$ and $[110]$ directions and $y$ is perpendicular to both $x$ and $z$. [Fig.~\ref{fig2}(a)] 
The hyperfine field $\bm{B}_{\rm hf}$ at the In sites located on a three-fold axis 
is written with the three Ni$^{2+}$ moments ($\bm{M}_1, \bm{M}_2$, $\bm{M}_3$) reproduced by the three-fold operation
and $\widehat{A}_{\rm hf}$ for each Ni$^{2+}$ moment rotated together by the three-fold rotation tensor $\widehat{R}_{z}$ as
\begin{align}
\bm{B}_{\rm hf} &= \widehat{R}_z^\mathsf{ T} \widehat{ A}_{\rm hf}\widehat{R}_z \bm{M}_1
+ \widehat{ A}_{\rm hf} \bm{M}_2+ 
\widehat{R}_z\widehat{ A}_{\rm hf}\widehat{R}_z^\mathsf{ T} \bm{M}_3 \notag\\ 
&= 3\widehat{ A}_{\rm tri} \bm{M} .\label{eq:Hhf}
\end{align}
In the paramagnetic state, where all Ni$^{2+}$ moments are polarized uniformly to the external field direction, that is, $\bm{M}_1 = \bm{M}_2 = \bm{M}_3$,
the total hyperfine coupling tensor becomes 
\begin{align}
\widehat{A}_{\rm tri} &= 
\left( 
\begin{array}{ccc}
\frac{1}{2}\left( A_{xx}+A_{yy} \right) & \frac{1}{2} \left( A_{xy}-A_{yx} \right) & 0 \\
-\frac{1}{2}\left( A_{xy}-A_{yx} \right) & \frac{1}{2}\left( A_{xx}+A_{yy} \right) & 0 \\
0 & 0 & A_{zz} \label{eq:Atri}
\end{array}
\right). 
\end{align}
Here, we note that the off-diagonal components between $xy$ and $z$ directions are canceled by the three-fold symmetry. 
On the other hand, an external field in the $y$ ($x$) direction contributes to the hyperfine field in the $x$ ($y$) direction through the off-diagonal components.
These hyperfine coupling constants are determined experimentally by measuring the Knight shift $K$ in corresponding external field directions.

\begin{figure}[tbp]
\begin{center}
\includegraphics[width=7.5cm]{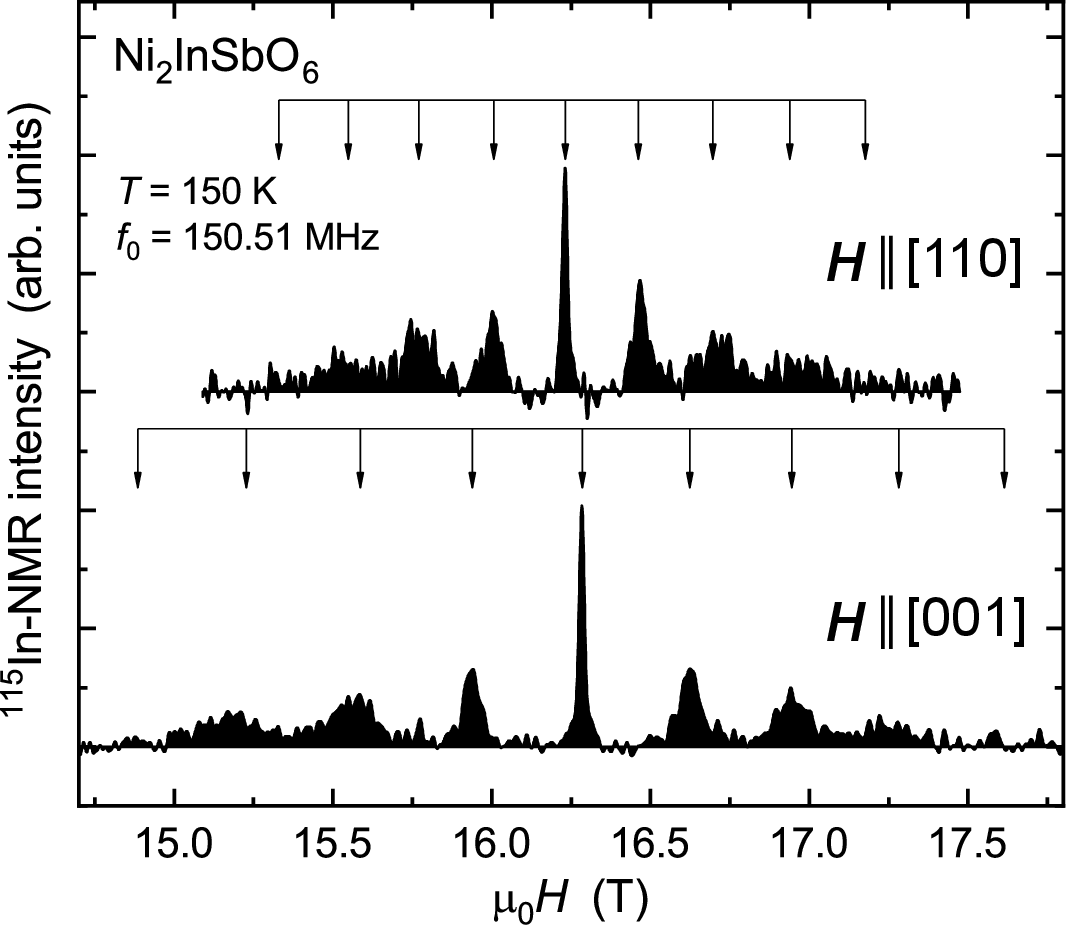}
\end{center}
\caption{
High temperature $^{115}$In-NMR spectra in fields along [110] (top) and [001] (bottom) directions. 
The measurement frequency is $150.51$ MHz for both directions. 
The $^{115}$In spectrum is split into 9 peaks by the electric quadrupolar interaction. 
The NQR frequency of 4.35 MHz was determined by the peak separation in $\bm{H} \parallel [001]$.
}
\label{fig3}
\end{figure}

Before measuring the temperature dependence of Knight shift,
we need to estimate the nuclear quadrupole resonance (NQR) frequency because the nuclear spin of In is $I=9/2$, 
and thus interacts with electric field gradient (EFG). 
Figure \ref{fig3} shows the $^{115}$In-NMR spectrum at temperature $T=150$ K $>T_N$ in fields applied parallel (bottom) and perpendicular (top) to the [001] axis. 
Together with the sharp central peak from $m = -1/2 \leftrightarrow 1/2$ transition broad satellite lines were observed. 
The resonant peaks from the highest order satellites were not clearly observed because of the spectral broadening originating from the local lattice distortion. 
By the three-fold local symmetry of the In site, the EFG tensor has axial symmetry around three-fold axis and the main principal axis of EFG is along the [001] axis. 
The NQR frequency along the [001] axis is determined by the peak separation as 4.35 MHz. 

\begin{figure}[tbp]
\begin{center}
\includegraphics[width=7.5cm]{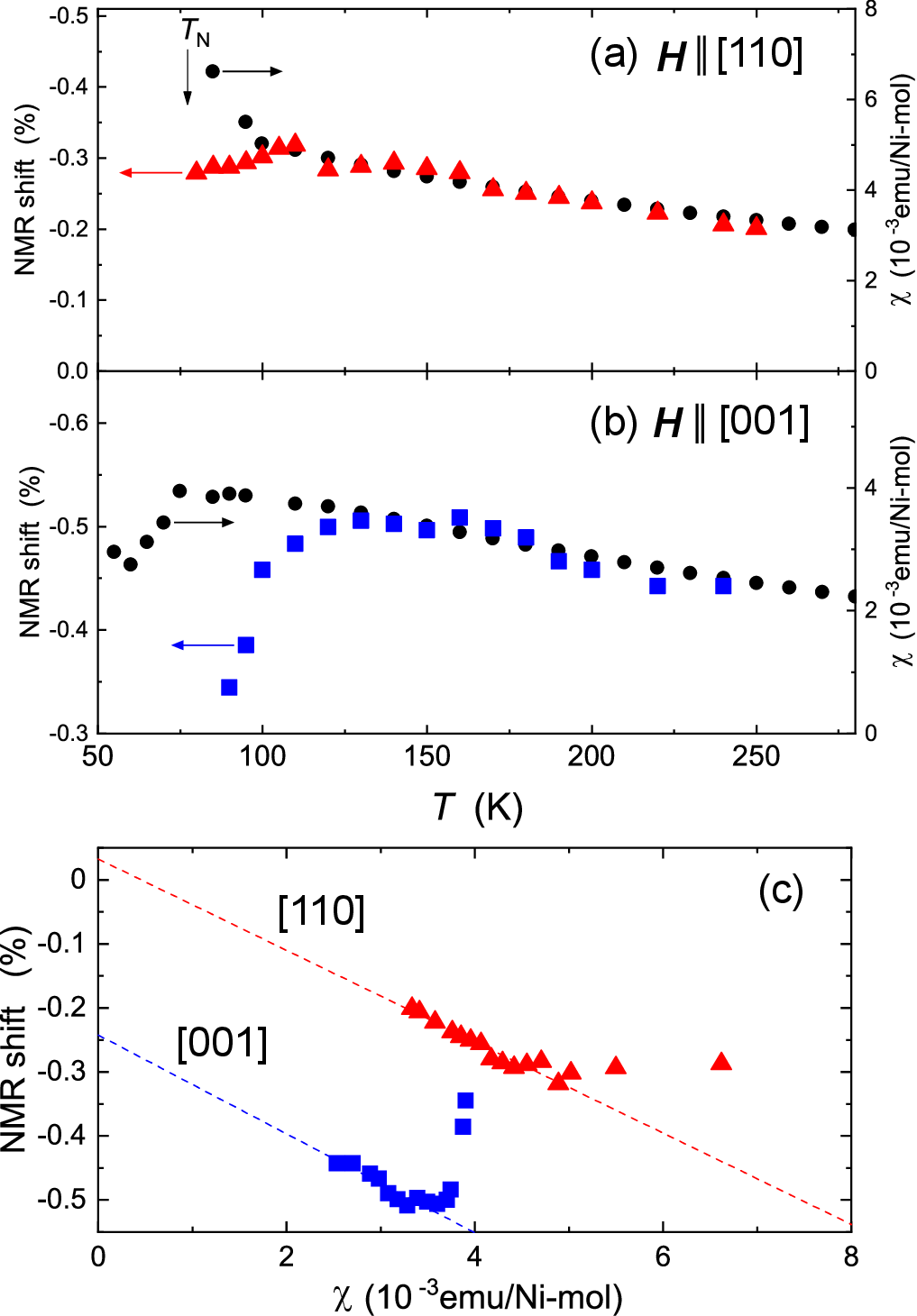}
\end{center}
\caption{
Temperature dependence of Knight shift plotted with bulk susceptibility for fields along (a) [110] and (b) [001]. 
Susceptibility is plotted against the right scale. 
(c) Knight shift is plotted against susceptibility using temperature as a implicit parameter. 
}
\label{fig4}
\end{figure}

We focus on the sharp central peak to measure the temperature dependence of Knight shift $K(T)$, 
which is shown in Figs.~\ref{fig4}(a) and (b) together with the bulk susceptibility $\chi(T)$ for corresponding field direction.  
In the paramagnetic state, $K$ is proportional to the bulk susceptibility, thus $K(T)=A_{\rm hf} \chi(T)$ for a certain field direction. 
Then by plotting $K(T)$ as a function of $\chi(T)$ using the temperature as an implicit parameter ($K-\chi$ plot), 
we can estimate $A_{\rm hf}$ from the slope of their linear relation.  
Figure \ref{fig4} (c) shows the $K-\chi$ plots for $\bm{H} \parallel [110]$ and $\bm{H} \parallel [001]$. 
The finite intercept $K_0$ originates from the quadrupolar shift.
Although the nuclear quadrupolar interaction does not modify $m=-1/2\leftrightarrow 1/2$ transition by the first order perturbation, the second order contribution gives a constant shift to the central peak.  
The linear relation between $K$ and $\chi$ was found at temperatures higher than 100 K for $\bm{H}\parallel [110]$ and 130 K for $\bm{H}\parallel [001]$. 
From the high-temperature linear relation the total coupling constants are obtained as $A^{\rm total}_{[110]}= -399$ mT/$\mu_B$ and $A^{\rm total}_{[001]} =-431$ mT/$\mu_B$. 
The dipole contributions are calculated from the crystal structure as $A^{\rm dip}_{[110]} = -43$ mT/$\mu_B$ and $A^{\rm dip}_{[001]} = 86$ mT/$\mu_B$. 
By subtracting the dipole contributions from $A_{\rm hf}^{\rm total}$, hyperfine coupling constants originating from the spin transfer interactions are estimated as $A^{\rm hf}_{[110]} = -356$ mT/$\mu_B$ and $A^{\rm hf}_{[001]} = -517$ mT/$\mu_B$.
These angle dependent hyperfine coupling constants are decomposed into the isotropic term $A_{\rm iso} =(2A^{\rm hf}_{[110]} + A^{\rm hf}_{[001]})/3 = -409$ mT/$\mu_B$  and 
the anisotropic term $A_{\rm ani} = (A^{\rm hf}_{[110]}-A^{\rm hf}_{[001]})/3=54$ mT/$\mu_B$.  
The anisotropic term originates from small population of In $5p$ electrons. 
Since $A_{\rm iso}$ is one order of magnitude larger than $A_{\rm ani}$, 
we ignore the anisotropic term for the analyses of NMR spectrum in the ordered state.

At lower temperatures near $T_N$ the Knight shift deviates from the bulk susceptibility, as shown in Figs.~\ref{fig4}(a) and (b), because of the development of short-range correlations, 
which disturbs the uniform polarization of paramagnetic spins and violates the linear relationship between uniform susceptibility and the microscopic internal fields at the In sites. 
The NMR intensity once disappears at $T_N$ because of the significant shortening of nuclear spin-spin relaxation time by critical fluctuations. 
The NMR intensity appears again when the critical fluctuations are suppressed at very low temperatures. 
In the ordered state below $T_N$, the Ni$^{2+}$ moments around the In sites create the spontaneous internal fields dominantly through the isotropic hyperfine interaction. 
The internal-field contribution from the direct dipole interaction was calculated to be less than $10$ mT in the magnetic structures studied here, 
which is much smaller than the internal fields created by the hyperfine interaction and observed actually from NMR measurement.
Therefore, to understand the magnetic structure in the ordered state, we need to estimate the hyperfine fields from each Ni$^{2+}$ moment around the In site. 

In the crystal structure of Ni$_2$InSbO$_6$, Ni$^{2+}$ ions occupy two crystallographic sites at Ni1:$(0,0,-0.0139)$ and Ni2:$(0,0,0.4798)$. \cite{ivanov-CM25} 
As shown in Fig.~\ref{fig2}(a), we find three Ni1 (Ni2) sites below (above) the target In nuclear spins. 
The three Ni sites around the In sites are reproduced by the three-fold operation. 
We label the nearest three Ni2 sites as Ni2-c to differentiate another Ni2 site (Ni2-f) located on the three-fold axis. 
The Ni2-f site has a significant contribution to the hyperfine interaction
because it is connected to the In site through three Ni-O-In paths. [Fig. \ref{fig2}(b)] 
Since the NiO$_6$ octahedron shares a face with the neighboring InO$_6$ octahedron, the coupling with the Ni2-f site should be the largest. 
Ni1 and Ni2-c sites are connected with the In site through two and one Ni-O-In paths, respectively.
The angle of the Ni-O-In path is close to 90 degrees for Ni2-f and Ni1 sites, while that for Ni2-c site is 120 degrees as listed in Table I. 
This structural feature suggests that the sign of hyperfine coupling constant for Ni2-c sites is opposite to those for Ni2-f and Ni1 sites.
As a result the total hyperfine coupling constant $A_{\rm hf} =A_{2f}+3A_{1}+ 3A_{2c}$ is partly canceled in the paramagnetic state. 

To roughly estimate the hyperfine coupling strength for each site, we refer to the exchange interaction calculated for Ni$_3$TeO$_6$ \cite{wu-IC49}, 
in which In and Sb are replaced with Ni (Ni3 sites) and Te, respectively, 
and thus the exchange interactions between Ni3 sites and Ni1, Ni2-f, and Ni2-c sites are mediated by the exchange paths similar to Ni-O-In bonds in Ni$_{2}$InSbO$_{6}$. (Table I)
We underline that the exchange interaction $J$ is negative only for Ni2-c sites, confirming the negative sign for $A_{2c}$. 
The relative values between these results weakly depend on the on-site Coulomb interaction. Assuming a similar ratio for the hyperfine coupling strength for Ni$_2$InSbO$_6$, 
we can estimate $(A_{1}, A_{2f}, A_{2c})=(+0.11, +0.56, -0.43)$ T/$\mu_B$.

\begin{table}
\caption{Exchange paths for Ni$_2$InSbO$_6$ (Ni$i$-O-In) and Ni$_3$TeO$_6$ (Ni$i$-O-Ni3)\cite{wu-IC49}. 
The last column shows the exchange interactions $J$ between Ni moments for each path. 
$J$ was calculated only for Ni$_3$TeO$_6$ \cite{wu-IC49}. }
\label{tab}
\begin{tabular}{|c|c|c|c|}\hline
Exchange path & Bond length (\AA) & Angle (deg.)& $J$ (meV) \\ \hline
\multicolumn{4}{|c|}{ Ni$_2$InSbO$_6$}\\ \hline
(Ni2-c)-O-In & 4.003 & 121.2 & --\\ \hline
Ni1-O1-In & 4.293 & 91.8 & \multirow{2}{*}{--} \\ \cline{1-3}
Ni1-O2-In & 4.287 & 92.0 & \\ \hline
(Ni2-f)-O-In & 4.252 & 86.3 & --\\ \hline
\multicolumn{4}{|c|}{ Ni$_3$TeO$_6$}\\ \hline
(Ni2-c)-O-Ni3 & 4.167 & 124.24 & -1.48 \\ \hline
Ni1-O1-Ni3 & 4.195 & 91.28 & \multirow{2}{*}{0.94} \\ \cline{1-3}
Ni1-O2-Ni3 & 4.186 & 91.55 & \\ \hline
(Ni2-f)-O-Ni3 & 4.273 & 81.07 & 4.55\\ \hline
\end{tabular}
\end{table}

\subsection{NMR spectrum at low magnetic fields}

\begin{figure}[tbp]
\begin{center}
\includegraphics[width=7.5cm]{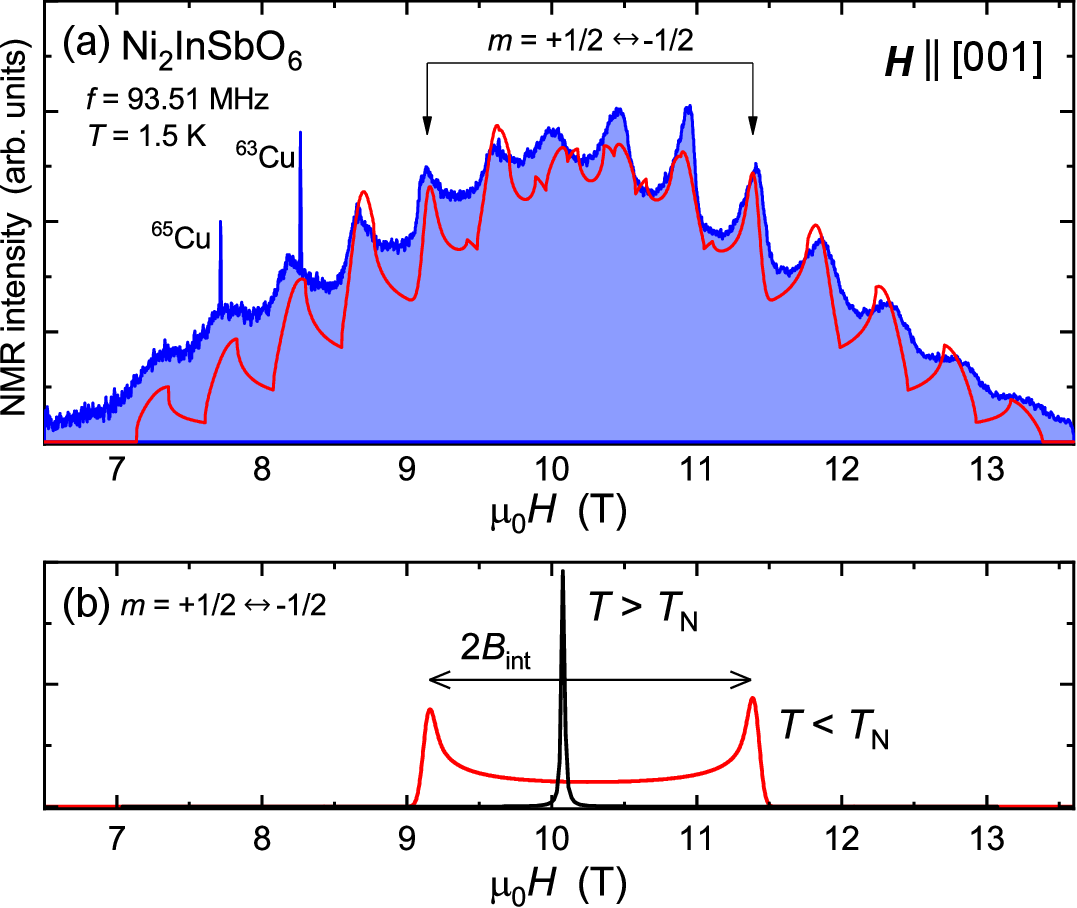}
\end{center}
\caption{
$^{115}$In-NMR spectrum in the LF phase at 1.5 K. 
Magnetic fields were applied in the [001] direction. 
Many peaks of saw-tooth shape are observed. 
(a) The experimental data (blue shadow) is consistently explained by the simulation (red solid line). 
The positions of two peaks split from the central $-1/2 \leftrightarrow 1/2$ transition as indicated by downward arrows. 
The sharp peaks of metallic $^{65}$Cu and $^{63}$Cu come from the rf coil. 
(b) The simulated spectral shape of only the central transition, $m=+1/2 \leftrightarrow -1/2$. 
The sharp peak in the paramagnetic state turns into a two-horn structure in the ordered state with spatial modulation of ordered moments. 
}
\label{fig5}
\end{figure}

With the microscopic parameters estimated above, we analyze the NMR spectrum in the ordered state, 
which is broadened magnetically by the spontaneous internal fields from the ordered Ni$^{2+}$ moments. 
Figure \ref{fig5}(a) shows the $^{115}$In-NMR spectrum for the LF phase measured in low magnetic fields applied in the [001] direction. 
In the LF phase the number of peaks becomes more than nine, which is the number of transitions for one nuclear spin $I = 9/2$, 
and each peak has an asymmetric shape like sawtooth. 
This spectral shape is explained by the spatially modulated internal fields typically observed for an incommensurate spin-density-wave state. 
The sinusoidal modulation of internal fields along the external field direction constructs a double-horn structure as shown in Fig.~\ref{fig5}(b). 
By imposing the same spectral broadening for all the transition lines and superposing these nine spectra, 
we successfully simulate the NMR spectrum in the ordered state, as shown by the red solid line in Fig.~\ref{fig5}(a). 

From the separation of two horns we can estimate the maximum value of the internal fields at the In site as $B_{\rm int} = 1.14$ T. 
This internal field is calculated by the following formula, 
\begin{align}\label{Bint}
\bm{B}_{\rm LF} = A_{2f} \bm{M}^{2f} + A_{1}\sum_{i} \bm{M}_i^{1}+ A_{2c}\sum_{i} \bm{M}_i^{2c}, 
\end{align}
where $\bm{M}$ is the ordered Ni$^{2+}$ moments forming the proper helical structure with propagation vector $\bm{k}=0.029b^{\star}$ and 
the index $i$ runs over three Ni sites around the three-fold axis. 
From the previous neutron diffraction study the size of the ordered moment was measured as $1.89 \mu_B$ 
and the Ni$^{2+}$ moments at Ni1 and Ni2 sites were suggested to be antiferromagnetically coupled. \cite{ivanov-CM25} 
In this spin configuration, the Ni$^{2+}$ moment at a position $(x,y,z)$ in the cartesian coordinate system defined in Fig. \ref{fig2}(a) is written as 
\begin{align}
\bm{M}(x,y,z) = 
M_0\left( 
\begin{array}{c}
\sin (2\pi ky + \phi_p) \\
0 \\
\cos (2\pi ky+ \phi_p) \\
\end{array}
\right).
\end{align}
Here, $M_0=1.89\mu_B, k=|\bm{k}|=0.029$, and $\phi_p$ is an arbitrary phase factor for Ni1 ($p=1$) and Ni2 ($p=2$) sites. 
For the AFM spin configuration between Ni1 and Ni2, which is realized by choosing $\phi_1 = \phi_2+\pi$, 
the maximum internal field in the $z$ direction is computed as 2.0 T using the isotropic hyperfine coupling constants estimated above. 
As this estimate is larger than the experimentally obtained internal field of $B_{\rm int} = 1.14$ T, 
we need to refine the Ni-site dependent hyperfine coupling constants. 
We note that the relative change between $A_{2f}$ and $A_{2c}$ is compensated in eq. (\ref{Bint}) by the constraint that 
the total hyperfine coupling constant must be consistent with the experimentally determined $A_{\rm iso} = -409$ mT/$\mu_B$. 
Therefore, we tuned the ratio between the hyperfine coupling constants to the Ni1 sites ($A_1$) and those for Ni2 sites ($A_{2f}$ and $A_{2c}$).
As the result, we obtained $(A_{1}, A_{2f}, A_{2c})=(+0.034, +0.38, -0.30)$ T/$\mu_B$. 
The relative decrease in $A_{1}$ with respect to $A_{2f}$ and $A_{2c}$ is consistent with the increase in the ratio for the corresponding bond lengths. 
For example, the ratio of Ni1-O1-Ni3 to (Ni2-c)-O-Ni3 bonds in Ni$_3$TeO$_6$ is 1.007, while the corresponding ratio of Ni1-O1-In to (Ni2-c)-O-In bonds in Ni$_2$InSbO$_6$ is 1.072.

\subsection{NMR spectra in high magnetic fields}

\begin{figure}[tbp]
\begin{center}
\includegraphics[width=8cm]{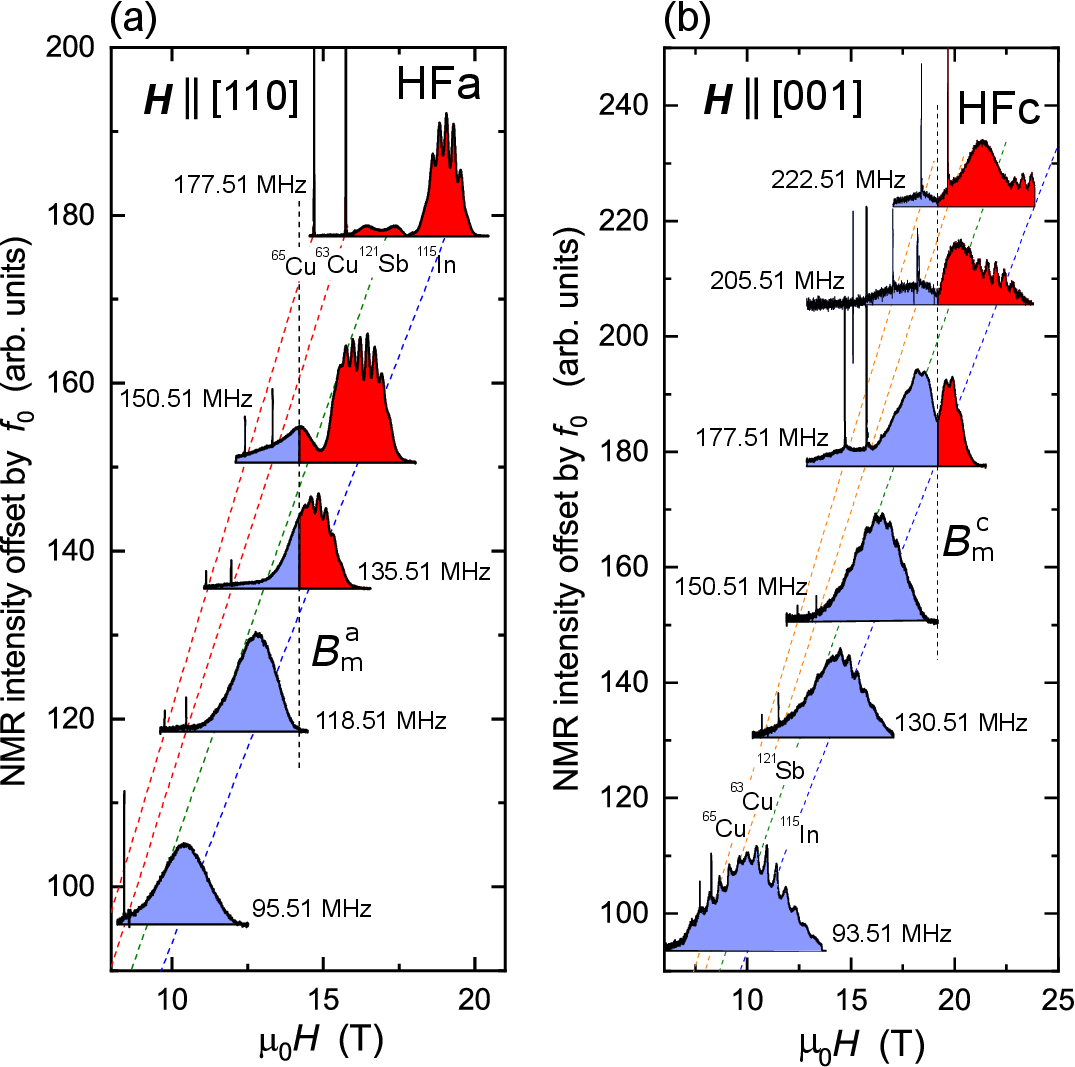}
\end{center}
\caption{
Field dependence of $^{115}$In-NMR spectra measured in the ordered states. 
The external fields were applied along (a) [110] and (b) [001] directions. 
The baseline was offset according to the measurement frequency for a better visibility. 
The reference fields for $^{63}$Cu, $^{65}$Cu, $^{121}$Sb, and $^{115}$In are indicated by dashed lines.
Change in the spectral shape was observed above the metamagnetic critical fields $B_m^a$ and $B_m^c$
}
\label{fig6}
\end{figure}

We measured the field dependence of the NMR spectra at 4 K for both fields parallel to [001] and [110] directions. 
Figure \ref{fig6} shows the field-sweep NMR spectra at several fixed frequencies. 
An apparent change in the spectral shape is observed at $\mu_0H=14.2$ T for $\bm{H}\parallel[110]$ and $\mu_0 H = 19.2$ T for $\bm{H} \parallel[001]$, 
both of which coincide with the field-induced anomalies detected by bulk measurements, \cite{araki-PRB102} 
suggesting that a change in magnetic structure affects the NMR spectral structures. 
In addition to the $^{115}$In-NMR signals, we observed the NMR spectra from the metallic $^{63}$Cu/$^{65}$Cu of the rf coil and from $^{121}$Sb in the sample. 
The reference fields at each measurement frequency are determined by the gyromagnetic ratio of each nuclear spins $\gamma$ as $f_{n} = \gamma_{n} H$ ($n=^{63}$Cu, $^{65}$Cu, $^{121}$Sb) 
and are represented by the dashed lines in Fig. \ref{fig6}. 
A broad $^{121}$Sb-NMR spectrum was observed only in the high field phases for both field orientations. 
This is because the optimal spin-echo pulse condition to obtain the largest $^{121}$Sb-NMR signal is different between LF and high-field phases. 
We optimized the pulse conditions to maximize the $^{115}$In-NMR signal intensity and focus only on the $^{115}$In-NMR spectra. 

To extract the information about the internal fields at the In site, 
we convert the horizontal axis of NMR spectrum at each frequency to $\Delta B = B_0- B_{\rm obs}$, 
where $B_0$ is a reference field determined by the measurement frequency $f_0$ as $B_0=f_0/\gamma$. 
For $\bm{H} \parallel [110]$, NMR spectrum at 118.51 MHz is in the LF phase and those above 131.51 MHz are in the HFa phase. 
The structureless broad spectra in the LF phase change to the multi-peak structure in the HFa phase. 
The spectral shape in the HFa phase is understood as the quadrupolar-split peaks with small magnetic broadening. 
The peak separation of $\sim 0.23$ T in the HFa phase is consistent with those observed at high temperature in Fig.~\ref{fig3}. 
The considerably small magnetic spectral width in the HFa phase suggests that the internal field at the In sites becomes spatially uniform after the metamagnetic transition. 
In contrast, the internal fields created in the LF phase with proper helical structure have spatial distribution, 
which leads to a spectrum broadening. 

In the LF phase, as the proper helical spin configuration does not have a spin component parallel to the field direction, \cite{araki-PRB102}
the internal field along the external field direction is produced by the off-diagonal terms of the hyperfine coupling tensor. 
We introduced a hyperfine coupling tensor $\widehat{A}_{\rm tri}$ in eq. (\ref{eq:Atri}) for the paramagnetic state, where all spins are aligned to the same direction. 
When spins take a proper helical structure with long wavelength, $\widehat{A}_{\rm tri}$ is modified 
because one of three spins is rotated by $9^{\circ}$ around the propagation direction. 
After a rotation of $\theta$ around $y$ axis, the magnetic moment $\bm{M}=(m_x, m_y, m_z)$ becomes 
\begin{align}
\bm{M}(\theta) &= 
\left( 
\begin{array}{c}
m_x\cos \theta - m_z \sin \theta \\
m_y \\
m_x\sin \theta + m_z \cos \theta \\
\end{array}
\right) \\
&\simeq \bm{M}(0) + 
\left( 
\begin{array}{c}
- m_z \\
0 \\
m_x \\
\end{array}
\right) \sin \theta. 
\end{align}
Here, we approximated $\cos 9^{\circ} = 0.99 \simeq 1$. 
The additional second term modifies $\widehat{A}_{\rm tri}$ by contributing to the off-diagonal ($xz$ and $zx$) components. 
Nevertheless, as these components are multiplied by a small factor of $\sin 9^{\circ} = 0.157$, we can treat the helical structure as a minor modification to the paramagnetic spin configuration at a microscopic length scale. 
Thus, the $z$ component of Ni$^{2+}$ moment does not produce any sizable internal fields in the $xy$ plane in the LF phase. 
When the external field is applied in the $y$ direction, the internal field in the $y$ direction is created mainly by the $x$ component of the Ni$^{2+}$ moments through the $xy$ component of $\widehat{A}_{\rm tri}$. 
The spatial modulation of $m_x=M_0\sin (2\pi ky+\phi)$ in the helical structure leads to the distribution of the internal fields, which then contributes to the large magnetic broadening of the NMR spectrum below $B_m^a$. 
The internal-field distribution from helical modulation is eliminated in the HFa phase with the canted AFM spin configuration, 
because $m_x$ is uniform over a large length scale when both Ni1 and Ni2 moments are fixed to a certain crystallographic orientation. [Fig.~\ref{fig8}(b)]
The field-induced NMR spectrum narrowing observed in the HFa phase evidences the canted AFM structure, which was proposed from the previous bulk measurements. \cite{araki-PRB102}

\begin{figure}[tbp]
\begin{center}
\includegraphics[width=7.5cm]{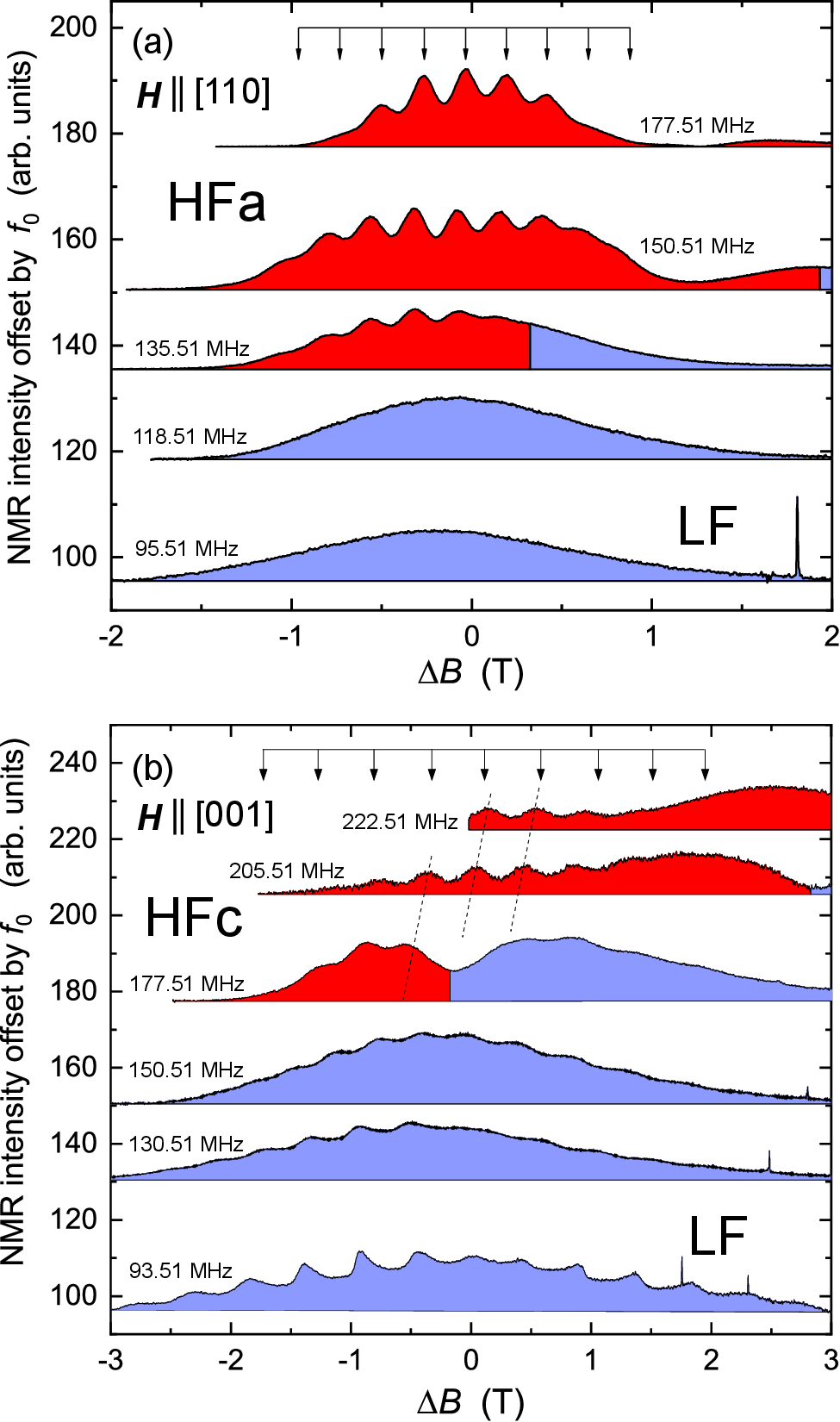}
\end{center}
\caption{
Internal-field representation of the $^{115}$In-NMR spectra. 
The horizontal axis $\Delta B$ is the shift from the reference field at each measurement frequency. 
The external field directions are (a) [110] and (b) [001]. 
Spectral width becomes narrow in HFa phase above $B_m^a$. 
In the field parallel to [001] direction, the saw-tooth shape in the LF phase disappears above $B_m^c$ and
quadrupolar-split NMR peaks are observed in the HFc phase. 
The peak position for the quadrupolar-split peaks are indicated by the downward arrow at the top of each figure. 
}
\label{fig7}
\end{figure}

In the magnetic fields parallel to [001], the reduction of spectral width was also observed in high-field (HFc) phase. 
The sawtooth structure in the LF phase is smeared at high fields and transforms to symmetric peaks above $19.2$ T. 
The separation of these symmetric peaks coincides with the quadrupolar splitting observed in the paramagnetic state (Fig. \ref{fig3}). 
The peak positions of quadrupolar-split 9-peaks for $\bm{H} \parallel [001]$ are indicated by downward arrows at the top of Fig. \ref{fig7}(b). 
As the NMR spectral shape is dominated by the quadrupolar splitting, we conclude that 
the magnetic broadening is much smaller than the quadrupolar interaction in the HFc phase as in the case of $\bm{H} \parallel [110]$. 
The full spectral shape for HFc phase was not observed because a part of the $^{115}$In-NMR spectrum at large internal fields ($\Delta B>2$ T) is overlapped with NMR signal coming from $^{121}$Sb nuclear spins in the sample. 
The other end ($\Delta B<0$ T at 222.51 MHz) was limited by the maximum magnetic fields currently accessible at IMR. 

\begin{figure}[tbp]
\begin{center}
\includegraphics[width=7.5cm]{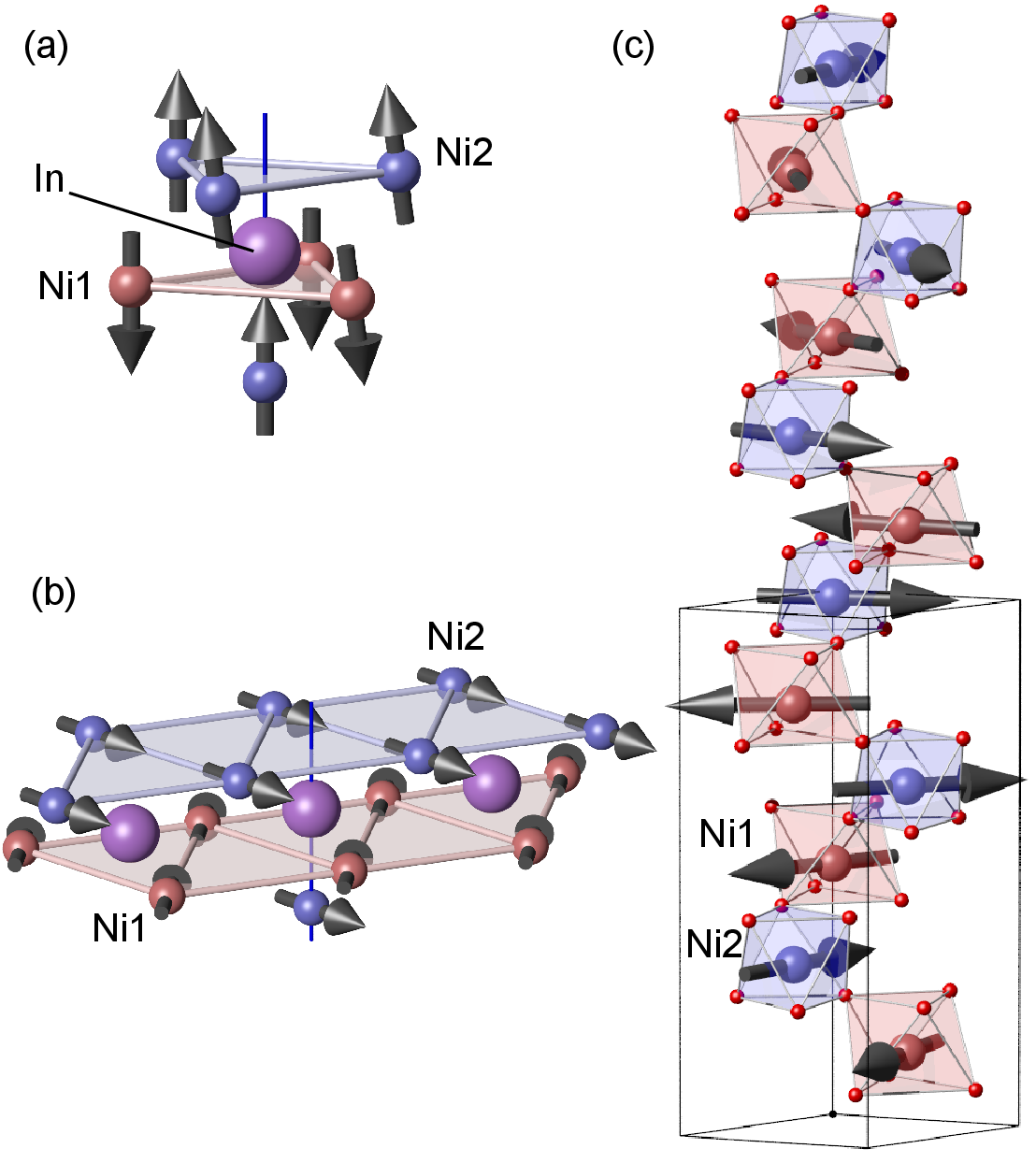}
\end{center}
\caption{
Magnetic structures suggested for (a) LF, (b) HFa and (c) HFc phases. 
(a) Ni spins in the triangular layer forms proper helical structure with a long wavelength. 
Ni1 and Ni2 spin magnetic moments are antiferromagnetically aligned. 
(b) In the HFa phase with fields applied in the [110] direction,  antiferomagnetically coupled Ni1 and Ni2 spin moments are canted toward the field direction. 
The spatial modulation of helical structure is removed by the ferromagnetic configuration of Ni spins on the same triangular layer. 
(c) In the HFc phase, propagation vector of the helical structure flops to the [001] direction.  
As the spins are nearly perpendicular to the external field direction, the internal fields along [001] direction is small. }
\label{fig8}
\end{figure}

The reduction of magnetic spectral width is consistently explained by the Q-flop transition at 19.2 T. 
At higher magnetic fields, the propagation vector points to the field direction $\parallel [001]$ and 
the Ni$^{2+}$ moments are aligned in the $(001)$ plane to form a proper helical structure along $[001]$ direction. [Fig.~\ref{fig8}(c)]
Here, we assume that Ni$^{2+}$ moments on the same plane are ferromagnetically aligned. 
This type of helical magnetic structure is realized in $Re$Be$_{13}$ series \cite{becker-MCLC125}. 
In this spin configuration, the hyperfine coupling tensor is written as eq. (\ref{eq:Atri}), as the Ni$^{2+}$ moments around the three-fold axis are aligned in parallel. 
Then, as the off-diagonal components between $z$ and $xy$ directions are zero, 
the ordered Ni$^{2+}$ moments aligned in the $(001)$ plane do not create any hyperfine fields along the [001] direction. 
As a consequence, the magnetic broadening is suppressed in the HFc phase. 
We note that in addition to the reduction of spectral broadening, the peak positions shift with fields, as indicated by tilted dashed line above 19 T in Fig.~\ref{fig7}(b).
This shift is interpreted as the partial spin polarization in the [001] direction after the Q-flop transition, 
which results in the conical spin structure. 

From recent neutron study a similar Q-flop transition was observed in a sister compound Ni$_{2}$ScSbO$_6$ by substituting a small amount of Co for Ni. \cite{ji-CC54}
The pristine Ni$_{2}$ScSbO$_6$ shows a helical magnetic ordering with long wavelength, as in the case of Ni$_{2}$InSbO$_{6}$. 
The magnetic propagation vector changes its orientation from in-plane to the $c$ direction in Ni$_{2-x}$Co$_x$ScSbO$_6$ with $x>0.5$. 
As the Co$^{2+}$ moments have a magnetic anisotropy stronger than Ni$^{2+}$ moments, 
the Q-flop transition can be attributed to the introduction of magnetic anisotropy. 
Assuming that the external magnetic fields also introduce the magnetic anisotropy, forcing magnetic moments to direct perpendicular to the external field direction, 
the HFc phase in Ni$_{2}$InSbO$_6$ would have the magnetic structure similar to those in Ni$_{2-x}$Co$_x$ScSbO$_6$. 
This similarity leads us to an intuitive understanding for the effect of magnetic fields in an AFM chiral magnet with competing interactions. 
Interestingly, the wavelength of helix in Ni$_{2-x}$Co$_x$ScSbO$_6$ is commensurate to the crystal structure and becomes short at small Co doping. 
In contrast to the Co doping, magnetic fields do not create any structural defects, and thus the Q-flop transition occurs in relatively uniform background. 
The wavelength and commensurability in HFc phase should be revealed to address the difference between the effects of magnetic field and Co doping. 
From the present NMR measurement, however, the wavelength in the HFc phase cannot be determined because NMR experiment detects only the local magnetic fields at the In sites 
but cannot measure the relative orientations between the neighboring Ni$^{2+}$ moments. 
The neutron diffraction measurement should be performed above $B_{m}^c=19.2$ T in future. 

\section{Conclusion}
We have investigated the magnetic structure of corundum-related antiferromagnet Ni$_2$InSbO$_6$ by means of $^{115}$In-NMR spectrum measurement. 
We observed the spectral broadening in the LF phase at low magnetic fields, which is ascribed to the spatial modulation of internal field induced by the helical magnetic structure with a long wavelength. 
The spectral width becomes narrower in high magnetic fields above the critical fields $B_m^{a}$ and $B_m^{c}$. 
Our microscopic analyses for the local magnetic fields at the In site lead us to determine the field-induced magnetic structure as a canted AFM for $\bm{H} \parallel [111]$ and Q-flopped helix for $\bm{H} \parallel [001]$. 
These magnetic structures are consistent with those proposed from the previous bulk measurements. \cite{araki-PRB102} 
By comparing the similar Q-flop transition in the Co-substituted Ni$_{2-x}$Co$_{x}$ScSbO$_6$, \cite{ji-CC54} 
we suggest that the high magnetic field modifies the magnetic structure by introducing the magnetic anisotropy. 
Present NMR results provide experimental evidence that high magnetic fields actually modify the helical magnetic structure even in the AFM chiral magnet. 
Microscopic identification of the field-induced magnetic structure establishes the field control of functional ME effects in the AFM chiral magnet  
and motivates technically challenging neutron diffraction measurements in high magnetic fields. 

\begin{acknowledgements}
The high-field NMR measurements were performed at the High Field Laboratory for Superconducting Materials, Institute for Materials Research, Tohoku University (Project Nos. 20H0601, 202110-HMCPB-0423)
This work was partially supported by JSPS KAKENHI (Grants Nos. 19H01832, 21H01035). 
\end{acknowledgements}

\end{document}